\documentclass[english,journal]{IEEEtran}
\usepackage[T1]{fontenc}
\usepackage[latin9]{inputenc}
\usepackage{units}
\usepackage{mathtools}
\usepackage{amsmath}
\usepackage{graphicx}

\makeatletter
\@ifundefined{date}{}{\date{}}
\usepackage{cite}
\usepackage[printonlyused]{acronym}
\acrodef{AWGN}{additive white Gaussian noise}
\acrodef{ASE}{amplified spontaneous emission}
\acrodef{QAM}{quadrature amplitude modulation}
\acrodef{PAM}{pulse amplitude modulation}
\acrodef{SE}{spectral efficiency}
\acrodef{SNR}{signal to noise ratio}
\acrodef{TX}{transmitter}
\acrodef{RX}{receiver}
\acrodef{BER}{bit error rate}
\acrodef{SER}{symbol error rate}
\acrodef{NFT}{nonlinear Fourier transform}
\acrodef{BNFT}{backward NFT}
\acrodef{FNFT}{forward NFT}
\acrodef{I-FNFT}{incremental FNFT}
\acrodef{DF-FNFT}{decision-feedback FNFT}
\acrodef{DF-BNFT}{decision-feedback BNFT}
\acrodef{NFDM}{nonlinear frequency-division multiplexing}
\acrodef{OFDM}{orthogonal frequency-division multiplexing}
\acrodef{NIS}{nonlinear inverse synthesis}
\acrodef{DAC}{digital-to-analog converter}
\acrodef{ADC}{analog-to-digital converter}
\acrodef{GVD}{group velocity dispersion}
\acrodef{SMF}{single mode fiber}
\acrodef{NCG}{Nystrom conjugate gradient}
\acrodef{B2B}{back-to-back}
\acrodef{EDC}{electronic dispersion compensation}
\acrodef{MAP}{maximum a posteriori probability}

\usepackage[linesnumbered,ruled,vlined]{algorithm2e}
\usepackage{algorithmic}

\makeatother

\usepackage{babel}
\begin{document}
\title{On the Nonlinear Shaping Gain with Probabilistic Shaping and Carrier
Phase Recovery}
\author{Stella Civelli\thanks{S.~Civelli, formerly with the Institute of Telecommunications, Computer
Engineering, and Photonics (TeCIP), Scuola Superiore Sant'Anna, Pisa,
Italy, is now with the Cnr-Istituto di Elettronica e di Ingegneria
dell\textquoteright Informazione e delle Telecomunicazioni, Pisa,
Italy, and with the National Laboratory of Photonic Networks, CNIT,
Pisa, Italy. E.~Forestieri and M.~Secondini are with the Institute
of Telecommunications, Computer Engineering, and Photonics (TeCIP),
Scuola Superiore Sant'Anna, Pisa, Italy, and with the National Laboratory
of Photonic Networks, CNIT, Pisa, Italy. E.~Parente is with the Institute
of Telecommunications, Computer Engineering, and Photonics (TeCIP),
Scuola Superiore Sant'Anna, Pisa, Italy. Email: stella.civelli@cnr.it.}, \IEEEmembership{Member, IEEE}, Emanuele Parente, Enrico Forestieri,
\IEEEmembership{Senior Member, IEEE}, and Marco Secondini, \IEEEmembership{Senior Member, IEEE}\thanks{This paper was presented in part at the European Conference on Optical
Communication (ECOC), Brusselles, Belgium, December 6-10, 2020. \cite{civelli2020interplayECOC}}}
\maketitle
\begin{abstract}
The performance of different probabilistic amplitude shaping (PAS)
techniques in the nonlinear regime is investigated, highlighting its
dependence on the PAS block length and the interaction with carrier
phase recovery (CPR). Different PAS implementations are considered,
based on different distribution matching (DM) techniques---namely,
sphere shaping, shell mapping with different number of shells, and
constant composition DM---and amplitude-to-symbol maps. When CPR
is not included, PAS with optimal block length provides a nonlinear
shaping gain with respect to a linearly optimized PAS (with infinite
block length); among the considered DM techniques, the largest gain
is obtained with sphere shaping. On the other hand, the nonlinear
shaping gain becomes smaller, or completely vanishes, when CPR is
included, meaning that in this case all the considered implementations
achieve a similar performance for a sufficiently long block length.
Similar results are obtained in different link configurations ($1\times180$\,km,
$15\times80$\,km, and $27\times80$\,km single-mode-fiber links),
and also including laser phase noise, except when in-line dispersion
compensation is used. Furthermore, we define a new metric, the nonlinear
phase noise (NPN) metric, which is based on the frequency resolved
logarithmic perturbation models and explains the interaction of CPR
and PAS. We show that the NPN metric is highly correlated with the
performance of the system. Our results suggest that, in general, the
optimization of PAS in the nonlinear regime should always account
for the presence of a CPR algorithm. In this case, the reduction of
the rate loss (obtained by using sphere shaping and increasing the
DM block length) turns out to be more important than the mitigation
of the nonlinear phase noise (obtained by using constant-energy DMs
and reducing the block length), the latter being already granted by
the CPR algorithm.
\end{abstract}


\section{Introduction\label{sec:Introduction}}

\IEEEPARstart{P}{robabilistic} amplitude shaping (PAS) has been recently
widely investigated as a way to improve the performance of an optical
fiber network. PAS allows to finely adapt the information rate to
the system requirements (channel signal-to-noise ratio (SNR) and forward
error correction (FEC) code) and to reduce the gap to the Shannon
limit in the linear regime \cite{bocherer2015bandwidth,buchali2016JLT,fehenberger2016JLT}.
The SNR gain---up to $1.53$\,dB for large constellation size \cite{kschischang1993optimal}---depends
on the particular implementation of PAS, handled by the distribution
matcher (DM). The DM maps $k$ independent input bits with uniform
distribution to $N$ output amplitudes with the desired Maxwell--Boltzmann
(MB) distribution---the optimal one in the linear regime. To do so,
the DM imposes some specific constraints (e.g., a constant composition
or a maximum energy) on the $N$ symbols of each block, which are
therefore \emph{correlated}. While a DM can be implemented in different
ways, its performance generally improves with the block length $N$.
In fact, the correlation between the symbols of each block decreases
when $N$ increases, allowing to encode more information per transmitted
symbol. For $N\rightarrow\infty$, the correlation vanishes and the
DM output looks like an i.i.d. source with MB distribution, yielding
the optimal PAS gain in the linear regime for a given rate and constellation
size \cite{kschischang1993optimal}.

Previous studies on PAS concerned the DM implementation and PAS performance
in the linear regime, aiming at reducing the rate loss---a useful
performance metric defined as the difference between the entropy of
the target MB distribution and the actual DM rate $k/N$---with reasonable
computational complexity, hardware requirements, and flexibility \cite{buchali2016JLT}.
For instance, sphere shaping (SS), implemented through the enumerative
sphere shaping (ESS) algorithm, provides the best performance for
a given block length \cite{gultekin2018Sphereshaping,gultekin2018approximate,gultekin2020probabilistic};
constant composition DM (CCDM), implemented by arithmetic coding,
is a simple and flexible technique to obtain the desired target distribution
\cite{schulte2016CCDM,fehenberger2018multiset}; hierarchical DM (Hi-DM)
is an effective approach to combine several short DMs (based, e.g.,
on simple look-up tables) to form a long DM with good performance
and low complexity \cite{yoshida2019hierarchicalDM,civelli2019hierarchicalOFC,civelli2020entropy,nadimigoki2021rateloss}.
In general, it was shown that increasing the block length $N$ of
the DM reduces the rate loss and improves the performance in the linear
regime, without any downside (but for the increased latency and difficulties
in the DM implementation). The interaction of carrier phase recovery
(CPR) algorithms and PAS was investigated in \cite{mello2018interplay},
considering the ideal MB distribution in the linear regime.

More recently, the performance of PAS in the nonlinear regime has
been studied \cite{fehenberger2016JLT}, and it was shown, firstly
for SS \cite{geller2016shaping} and later for CCDM \cite{amari2019introducing,fehenberger2020mitigating},
that increasing the block length at will is not beneficial. Indeed,
the constraints induced by the DM on the $N$ symbols of each block,
besides reducing the rate of the source causing a rate loss, usually
reduce also the intensity fluctuations on the signal, hence reducing
the amount of nonlinear interference generated by each channel and
yielding an additional \emph{nonlinear shaping gain}. In this case,
the correlation induced by the DM is beneficial, so that the nonlinear
shaping gain decreases as $N$ increases, vanishing for $N\to\infty$.
Therefore, there is an optimal block length that maximizes the shaping
gain by providing the best trade-off between linear and nonlinear
gain. The nonlinear interference due to DM was analyzed in \cite{wu2021temporal}
for the CCDM, while the kurtosis-limited sphere shaping, which selects
the sequences with minimum energy and low kurtosis, showed superior
performance in the nonlinear regime with respect to equivalent-length
ESS in a single-span scenario but not for a multi-span link \cite{gultekin2021kurtosis}.
Furthermore, it was shown that the nonlinear shaping gain improves
by properly packing shaped sequences in time and frequency \cite{cho2021shaping}.

However, it is also known that a good part of the inter-channel nonlinear
interference generated by intensity fluctuations (which are reduced
by a short-block-length DM, as explained above) manifests as correlated
phase noise, which can be mitigated also by a properly optimized CPR
algorithm \cite{secondini2012analytical,Dar2013:opex,dar_JLT2017_nonlinear,yankov:JLT2015,secondini2019JLT}.
Unfortunately, a preliminary study on the interaction between the
nonlinear shaping gain and CPR algorithms showed that the gain provided
by the two techniques is very similar and does not add up \cite{civelli2020interplayECOC}.
Similar conclusions were drawn by an analytical study about the interaction
of CPR and CCDM for cross phase modulation \cite{rafie2022constant}.
This effect is particularly relevant from a system design perspective,
since a carrier recovery algorithm is always included in practical
systems, meaning that the nonlinear shaping gain observed in simulations
in the absence of a carrier recovery algorithm, might in fact disappear
(or be drastically reduced) in realistic systems. In this work, we
extend the analysis in \cite{civelli2020interplayECOC} to assess
the interaction between CPR and PAS in terms of nonlinearity mitigation
in a wavelength-division multiplexing (WDM) scenario. This is done
by including the laser phase noise in the system, highlighting the
performance of different PAS and DM implementations, considering different
scenarios, and introducing a new performance metric to study and predict
this interaction.
\begin{figure}[tp]
\centering

\includegraphics[width=1\columnwidth]{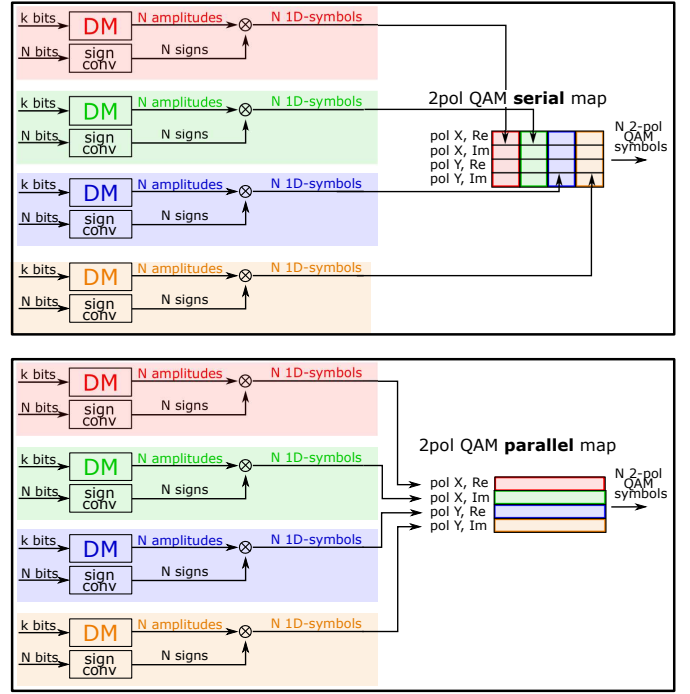}\caption{\label{fig:setup_mapping}Different mappings to generate $N$ 2-pol
QAM symbols from $4N$ amplitudes generated by four instances of the
same DM and $4N$ i.i.d. bits used for the signs.}
\end{figure}

\section{Probabilistic Amplitude Shaping\label{sec:Probabilistic-amplitude-shaping}}

\begin{figure}[tp]
\centering

\includegraphics[width=1\columnwidth]{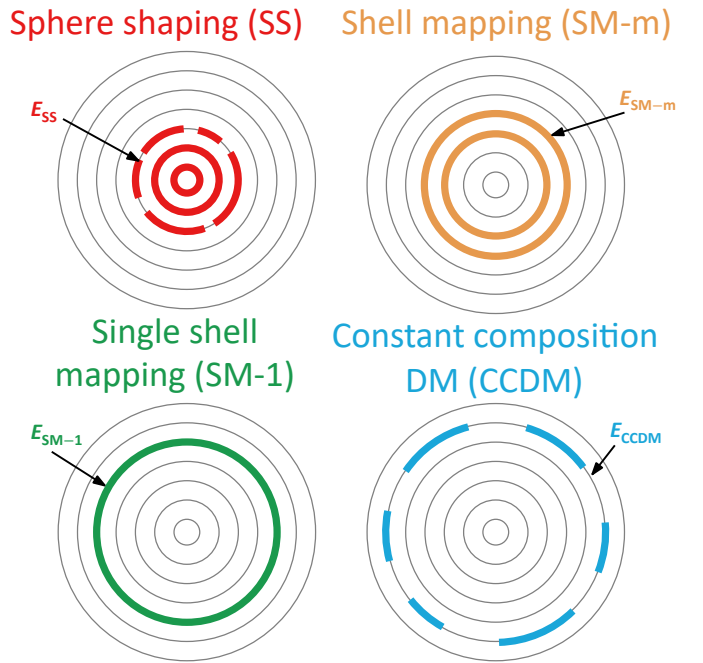}\caption{\label{fig:setup_DM}2D energy description of different DM techniques.
Black thin circles represent possible energy levels.}
\end{figure}

PAS is implemented at the transmitter by using four identical DMs.
Each DM maps $k$ uniformly distributed bits to $N$ shaped amplitudes,
$A_{1},\ldots,A_{N}$. The $4N$ amplitudes are then combined with
$4N$ signs (obtained from other $4N$ uniform bits) and mapped to
the four components of $N$ dual-polarization QAM symbols (i.e., 4D
symbols).\footnote{The reverse concatenation with the FEC is irrelevant to this description
and is, therefore, omitted.} The amplitude-to-symbol mapping can be done in different manners.
Here, we consider the two maps sketched in Fig.\ \ref{fig:setup_mapping},
referred to as \emph{serial map} and \emph{parallel map}. On the one
hand, the serial map maps the $N$ amplitudes generated by the first
DM to the four components of the first $N/4$ 4D symbols, the $N$
amplitudes generated by the second DM to the four components of the
next $N/4$ 4D symbols, and so on. On the other hand, the parallel
map maps the $N$ amplitudes of the first DM to the first component
of the $N$ 4D symbols, the $N$ amplitudes of the second DM to the
second component of the $N$ 4D symbols, and so on. While the serial
map induces a stronger correlation (the four components are correlated)
on a shorter block of $N/4$ adjacent symbols, the parallel map induces
a weaker correlation (the four components are independent) on a longer
block of $N$ adjacent symbols. The two maps are equivalent in the
linear regime but, as we will see in the following, they provide different
performance in the nonlinear regime.

For the PAS implementation, we consider different DM techniques: SS,
shell mapping (SM), and CCDM, as described below. The energy distribution
of the methods is qualitatively depicted in Fig.~\ref{fig:setup_DM}.\footnote{While in our manuscript we think of SS as a type of DM, a different
perspective is given in \cite{gultekin2020probabilistic}, where the
SS is proposed as an \emph{indirect} method to implement PAS, as opposed
to the \emph{direct} methods implemented by a DM. This difference
is, however, only a matter of definition and does not affect the practical
realization of PAS.}

SS maps $k$ bits to the $2^{k}$ lowest-energy sequences of $N$
amplitudes. Thus, by representing each sequence as a point in an N-dimensional
space, all the sequences must lie within the smallest possible sphere
that contains at least $2^{k}$ sequences. The map covers all the
sequences \emph{inside} the sphere and some of the sequences \emph{on}
the sphere, as shown qualitatively in Fig.~\ref{fig:setup_DM}. For
a given block length $N$ and constellation size, SS maximizes the
source rate for a given average energy, yielding the best performance
in the linear regime. The average energy of the sequences is $E_{\text{SS}}$.
In this manuscript, SS is implemented using the ESS algorithm \cite{willems1993pragmatic,gultekin2018Sphereshaping,gultekin_TWC_2019_enumerative},
resorting to the double-trellis trick proposed in \cite{civelli2020entropy}
and studied in \cite{chen2021squeezing} to obtain optimal performance;
however, this is not the only way to implement SS \cite{gultekin2020probabilistic}
and a simple look-up-table could be used for short-block-lengths.

SM maps $k$ bits to the $2^{k}$ lowest-energy sequences, with the
additional constraint that at most $m$ energy levels (shells) can
be occupied---indicated as SM-$m$. Thus, SM-1 uses only sequences
that lie in a single shell and, therefore, have the same energy; SM-2
uses two adjacent shells; and so on. When the number of shells increases,
SM-$m$ tends to SS. The energy of the sequences covered by SM-$m$
is limited by a maximum and a minimum value, with $E_{\text{SM-}m}$
being the average energy. Also SM is implemented by using the ESS
algorithm; the recently proposed band-ESS can also be used \cite{gultekin2022mitigating}.
In the following, we consider only two extreme cases: the single-shell
case, denoted as SM-1, and the case with the maximum number of shells
(but lower than SS), denoted as SM-max. The latter is obtained by
adding one higher-energy shell to those used by SS and removing all
the innermost ones that are no longer needed (the number of shells
$m$ varies in this case).

CCDM maps $k$ bits on $2^{k}$ amplitude sequences with the same
composition, i.e., permutations of the same sequence \cite{schulte2016CCDM,shapecomm_webdm}.
The composition is determined by the desired target distribution.
Since the sequences have the same composition, they also have the
same energy $E_{\text{CCDM}}$ and lie in a single shell, as in the
SM-1 case.  

In the linear regime, for a given block length $N$ and constellation
size, the PAS performance depends on the average energy $E_{\mathrm{\text{DM}}}$
of the $2^{k}$ sequences used by the considered DM. It is simple
to verify that $E_{\text{CCDM}}\geq E_{\text{SM-1}}\geq E_{\text{SM-2}}\geq\dots\ge E_{\mathrm{\text{SM}-max}}\geq E_{\text{SS}}$.
Thus, the best performance is obtained with SS, then with SM-$m$
(the performance decreasing with decreasing $m$), and eventually
with CCDM. As the block length $N$ increases, all the mentioned methods
approach an i.i.d. source with MB distribution \cite{kschischang1993optimal},
which yields the ultimate linear shaping gain (and zero rate and energy
losses). For a given block length and constellation size, the rate
loss, which is a very common performance metric for DMs in the linear
regime, follows the same ranking indicated by the average energy,
as shown in several recent publications including \cite{gultekin_TWC_2019_enumerative,amari2019introducing}.

On the other hand, in the nonlinear regime, the capacity-achieving
distribution and, consequently, the optimal DM are unknown. However,
some useful design guidelines can be obtained from approximated models
or observations. For instance, it has been shown that the amount of
nonlinear interference generated by a propagating signal depends not
only on its average power (second-order moment), but also on its fourth-order
moment (when symbols are i.i.d.) \cite{Mecozzi:JLT0612,Secondini:JLT2013-AIR,Dar2013:opex,Carena:OPEX14}
or, more in general, on the variations of the instantaneous power
over a finite temporal window \cite{wu2021temporal,wu2021EEDI,cho2021shaping}.
In this case, the use of a shorter block length $N$ is expected to
be beneficial, as it introduces a constraint on the energy of each
block of $N$ symbols. The constraint is stronger for CCDM and SM-1
(for which the energy of each sequence if constant), and becomes weaker
for SM-$m$ as $m$ increases (since the energy of each sequence may
take $m$ different values). Therefore, as opposed to the ranking
defined in terms of linear performance, we expect CCDM and SM-1 to
provide the most effective nonlinearity mitigation, followed by SM-2,
SM-3, and so on, while SS should be the least effective. As a result,
the optimization of both the DM type and its block length should aim
to obtain the best trade-off between two conflicting objectives: the
reduction of the rate loss (linear shaping gain) and the reduction
of the intensity fluctuations (nonlinear shaping gain). This will
be investigated in Section~\ref{sec:Conclusion}.

To study the behavior of different DMs as a function of the block
length $N$, we resort to a trick to emulate SS and SM-$m$ for $N>512$,
when the rate loss of the two methods is very small, but the computation
of the required trellis structures becomes nearly unfeasible. In this
case, we concatenate the amplitudes generated by $N/512$ independent
uses of a DM of block length $512$, followed by an interleaver of
length $N$. In this manner, we emulate the correlation induced by
a DM of block length $N$, while achieving almost the same rate loss.

\section{Nonlinear Phase Noise Metric\label{sec:Nonlinear-phase-noise}}

Different fiber nonlinearity models agree that a relevant portion
of nonlinear interference---in particular that generated by the intensity
fluctuations of the signal---manifests as phase noise \cite{Mecozzi:JLT0612,secondini2012analytical,Secondini:JLT2013-AIR,Dar2013:opex}.
For instance, the frequency-resolved logarithmic perturbation (FRLP)
model describes nonlinear interference as a frequency-dependent phase
noise that can be expressed as a quadratic form of the symbols transmitted
in a certain time window around the considered time on the (self-
or cross-) interfering channels \cite{secondini2012analytical,Secondini:JLT2013-AIR}.\footnote{Though the window formally extend over an infinite time, the coefficients
of the quadratic form rapidly decays outside a finite time window
determined by the walk-off time between the frequency components involved
in the interference process.} For i.i.d. symbols, the variance of this phase noise depends on the
kurtosis of the symbols, so that non-constant-envelope modulations
(e.g., QAM or Gaussian modulation) cause a stronger NLI than constant-envelope
modulations (e.g., PSK) \cite{Secondini:JLT2013-AIR}. However, the
nonlinear phase noise generated by non-constant-envelope modulations
is strongly correlated in time, so that it can be mitigated by a suitable
carrier-phase recovery algorithm, after which its variance practically
reduces to that of constant-envelope modulation \cite{Sec:PTL14}.
Here, we extend the analysis to the case of correlated symbols, such
as those generated by the PAS schemes described in Section~\ref{sec:Probabilistic-amplitude-shaping},
with the aim of finding a suitable metric to predict the dependence
of the generated NLI on the PAS block length, accounting for the possible
presence of a carrier-phase recovery algorithm. With respect to \cite{Secondini:JLT2013-AIR},
since removing the i.i.d. assumption significantly complicates the
analysis \cite{liga2020extending}, we further simplify the FRLP
model to obtain a nonlinear phase noise model that depends only on
signal intensity, and resort to a numerical approach for the computation
of the variance.

The simplified nonlinear phase noise model is derived from the FRLP
model \cite{Secondini:JLT2013-AIR} by following the same approach
used in \cite{secondini2014enhanced,secondini_PNET2016,civelli2021coupled,civelli2021ISWCS2021}
to develop the enhanced split-step Fourier method (ESSFM) and the
coupled-channel ESSFM (CCESSFM) algorithms for DBP. By considering
only the terms on the diagonal of the quadratic form in \cite[eq. (18)]{Secondini:JLT2013-AIR},
neglecting (or averaging out) their dependence on frequency, and accounting
for the contributions of the two polarizations of each interfering
channel, we eventually obtain a simple phase noise term that depends
on the intensity variations of all the interfering channels. Specifically,
considering $M$ WDM channels, denoting by $x_{i}[k]$ and $y_{i}[k]$
the normalized $k$th symbols transmitted on the two polarizations
of the $i$th channel, the corresponding output samples after dispersion
compensation can be expressed as $x_{i}[k]\exp(-j\theta_{i}[k])$
and $y_{i}[k]\exp(-j\theta_{i}[k])$, where 
\begin{equation}
\theta_{i}[k]=\sum_{\ell=1}^{M}\bar{\phi}_{i\ell}\sum_{\mathclap{m=-N_{c}}}^{N_{c}}C_{\ell-i}[m]\left(|x_{\ell}[k+m]|^{2}+|y_{\ell}[k+m]|^{2}\right)\label{eq:CCESSFM_singlepol}
\end{equation}
is the overall nonlinear phase rotation \cite{civelli2021ISWCS2021}.
In (\ref{eq:CCESSFM_singlepol}), $C_{\ell-i}[m]$ ($m=-N_{c},\dots,N_{c}$)
are $2N_{c}+1$ real coefficients accounting for the interaction of
dispersion and nonlinearity induced by channel $\ell$ over channel
$i$; $\bar{\phi}_{i\ell}=(3/2-\delta_{i\ell}/2)\gamma\int_{0}^{L}P_{\ell}(\zeta)d\zeta$
is the average nonlinear phase rotation induced by channel $\ell$
over channel $i$, with $P_{\ell}(\zeta)$ the power of channel $\ell$
at distance $\zeta$, $L$ the length of the link, $\gamma$ the nonlinear
coefficient of the fiber, and $\delta_{i\ell}$ the Kronecker delta.
The coefficients can be evaluated analytically from \cite[eq. (19)]{Secondini:JLT2013-AIR}\footnote{In this case, we consider only the diagonal terms of the quadratic
form, neglect their dependence on frequency, assume an ideal sinc
pulse shape, and adopt a different normalization} as

\begin{equation}
{\displaystyle C_{n}[m]}=T^{2}\int_{\frac{2n-1}{2T}}^{\frac{2n+1}{2T}}\int_{\frac{2n-1}{2T}}^{\frac{2n+1}{2T}}K(\mu,\nu)e^{-j2\pi(\mu-\nu)mT}\,\text{d}\mu\text{d}\nu\label{eq:Canal}
\end{equation}
where $T$ is the symbol time and $K(\mu,\nu)$ is a function that
depends on the link characteristics and accounts for the nonlinear
interaction efficiency of different frequency components. Considering
a dispersion-unmanaged link made of $N_{s}$ identical fiber spans
of length $L_{s}$, with attenuation coefficient $\alpha,$ dispersion
parameter $\beta_{2}$, nonlinear coefficient $\gamma$, and ideal
dispersion compensation at the end of the link, we have
\begin{align}
K(\mu,\nu) & =\frac{\exp\left[-\alpha+j4\pi^{2}\beta_{2}\nu(\nu-\mu)L_{s}\right]-1}{-\alpha+j4\pi^{2}\beta_{2}\nu(\nu-\mu)}\nonumber \\
 & \times\frac{\exp\left[j4\pi^{2}\beta_{2}\nu(\nu-\mu)N_{s}L_{s}\right]-1}{\exp\left[j4\pi^{2}\beta_{2}\nu(\nu-\mu)L_{s}\right]-1}\frac{\alpha}{N_{s}(1-e^{-\alpha L_{s}})}
\end{align}

Interestingly, there is a clear similarity between the nonlinear phase
rotation in (\ref{eq:CCESSFM_singlepol}) and the weighted sum of
symbol energies used in \cite{wu2021temporal,wu2021EEDI} to define
the energy dispersion index (EDI) and the exponentially-weighted EDI
(EEDI) and predict the nonlinear shaping gain that occurs in a PAS
system using CCDM. For example, the EEDI is defined as \cite{wu2021EEDI}
\begin{equation}
\text{EEDI}=\frac{\text{Var}(G^{\lambda}[k])}{E(G^{\lambda}[k])}\label{eq:EEDI}
\end{equation}
where
\begin{equation}
G^{\lambda}[k]=\sum_{m=-\infty}^{+\infty}\lambda^{|m|}|x_{i}[k+m]|^{2}\label{eq:weighted_sum_energies}
\end{equation}
and $0\leq\lambda\leq1$ is a forgetting factor.\footnote{A similar definition holds for the EDI, with the only difference of
considering a finite sum of $W$ elements with $\lambda=1$.} This similarity provides a physical explanation of why EDI and EEDI
are good predictors of the nonlinear shaping gain (when they are small,
the signal is affected by less nonlinear phase noise). The main difference
between (\ref{eq:weighted_sum_energies}) and (\ref{eq:CCESSFM_singlepol})
is that the coefficients in (\ref{eq:CCESSFM_singlepol}) depend on
the link characteristics and can be obtained analytically, while the
parameter $\lambda$ in (\ref{eq:weighted_sum_energies}) (or the
window length $W$ in the EDI) is optimized a posteriori, through
extensive simulations, to maximize the correlation with the system
performance. Moreover, (\ref{eq:CCESSFM_singlepol}) accounts for
both polarizations and includes the interfering channels.\footnote{The recently proposed lowpass filtered symbol-amplitude sequences
(LSAS) metric also allows to account for inter-polarization and inter-channel
effects \cite{askari2022ecoc}.} Therefore, we propose to replace the EDI or EEDI with the variance
of (\ref{eq:CCESSFM_singlepol}) as a predictor of the nonlinear shaping
gain of PAS systems. This solution avoids the use of extensive simulations,
poses performance prediction on a more physical ground---relating
it to the amount of nonlinear phase noise accumulated during propagation---and,
as shown below, allows to easily account for the impact of CPR.

In order to account for the randomness of the carrier phase and its
temporal variations due to laser phase noise, coherent optical receivers
usually include a CPR algorithm. Practical algorithms, though not
specifically designed for this purpose, can partly mitigate also the
nonlinear phase noise in (\ref{eq:CCESSFM_singlepol}), reducing its
variance. Of course, the amount of nonlinearity that can be mitigated
depends on the specific algorithm and on the width of the time window
over which the carrier phase is estimated (or any other parameter
playing an analogous role). As with the conventional phase noise due
to lasers, a longer time window allows to average out more effectively
the impact of additive white Gaussian noise on the estimate, but reduces
the ability to track fast changes of the phase.

The above considerations have two important consequences. First, when
estimating the effectiveness of PAS as a nonlinear mitigation strategy,
the presence of CPR should be accounted for. Second, the optimization
of the PAS block length $N$ and of the CPR half time window $N_{\mathrm{\text{CPR}}}$
are intertwined: they depend on each other and should be done jointly,
accounting for the link configuration, the signal-to-noise ratio,
and the laser linewidth.

With this in mind we define the nonlinear phase noise (NPN) metric
(for the generic $i$th channel) as
\begin{equation}
\sigma_{\theta}^{2}=\text{Var}(\theta_{i}[k]-\hat{\theta}_{i}[k])+\sigma_{\xi}^{2}\label{eq:RNPN-metric}
\end{equation}
where
\begin{equation}
\hat{\theta}_{i}[k]=\frac{\sum_{m=-N_{\text{CPR}}}^{N_{\text{CPR}}}\theta_{i}[m+k]}{2N_{\text{CPR}}+1}.\label{eq:theta_estimated}
\end{equation}
is the noiseless estimate of the nonlinear phase rotation at time
$k$ provided by the CPR, which we assume to equal the average nonlinear
phase noise measured over a window of $2N_{\text{CPR}}+1$ symbols
around the $k$th symbol, and $\sigma_{\xi}^{2}$ is the variance
of the noise affecting the CPR estimate. In general, the latter depends
on the specific CPR algorithm, on its block length (or other equivalent
parameter), and on the signal-to-noise ratio (SNR). Here, for the
sake of simplicity, we assume
\begin{equation}
\sigma_{\xi}^{2}=\frac{(2E_{s}/N_{0})^{-1}/(2N_{\text{CPR}}+1)}{e_{\text{\ensuremath{\mathrm{CPR}}}}}\label{eq:CPRnoise}
\end{equation}
where the numerator is the Cram�r--Rao lower bound, and $e_{\text{\text{\ensuremath{\mathrm{CPR}}}}}\le1$
is a coefficient that measures the efficiency of the CPR algorithm
with respect to the bound \cite{pfau2009BPS}. In principle, different
definitions of (\ref{eq:theta_estimated}) and (\ref{eq:CPRnoise})
could be employed to account more precisely for the actual behavior
of a specific CPR algorithm, though this is outside the scope of this
work.

\begin{figure*}[tp]
\centering

\includegraphics[width=1.5\columnwidth]{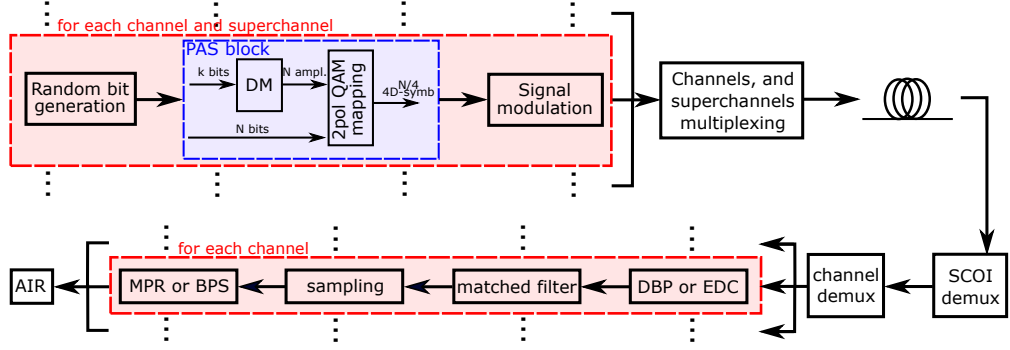}\caption{\label{fig:setup}System setup: WDM modulation with PAS block, fiber
propagation, and AIR evaluation after channel demultiplexing, EDC
or DBP and MPR or BPS. }
\end{figure*}

\section{Simulation Setup\label{sec:simulation-setup}}

The system setup is sketched in Fig.~\ref{fig:setup}, and is the
same considered in \cite{civelli2020interplayECOC}. A stream of uniformly
distributed bits---representing the information bits after FEC encoding---feeds
the PAS block (see Section~\ref{sec:Probabilistic-amplitude-shaping}),
which maps the bits to symbols of a dual polarization $256$ quadrature
amplitude modulated (QAM) constellation with rate $6$ bits/symbol/pol.
Using a root raised cosine (RRC) pulse with rolloff $0.1$ and baud
rate $R_{s}=41.67$\,GBd, the signals corresponding to $4$ adjacent
channels are multiplexed in a single superchannel, the superchannel
of interest (SCOI), with $75$\,GHz spacing. Two additional superchannels,
with the same properties of the SCOI, are also multiplexed, such that
12 evenly spaced channels are transmitted over an overall bandwidth
of $900$\,GHz. The generated WDM waveform is launched into the link,
composed of several spans of $80$\,km single mode fiber (SMF) with
dispersion $D=17\thinspace\text{ps}/\text{nm}/\text{km}$, Kerr parameter
$\gamma=1.3\thinspace\text{W}^{-1}\text{km}^{-1}$, and attenuation
$\alpha_{\text{dB}}=0.2\thinspace\text{dB}/\text{km}$. After each
span, an erbium-doped fiber amplifier (EDFA) with a noise figure of
$5$\,dB compensates for loss. At the end of the link, the side superchannels
are filtered out, and the $4$ channels of the SCOI are demultiplexed.
The transmitter and receiver laser linewidth $\Delta\nu$ are set
to either $0$ or $100$~kHz. Each channel undergoes: (i) either
ideal digital back propagation (DBP) or electronic dispersion compensation
(EDC), (ii) matched filtering, (iii) sampling at symbol time, and
(iv) CPR. Finally, the average achievable information rate (AIR)
of the $4$ channels of the SCOI is evaluated, representing the average
information per symbol that can be reliably transmitted on each polarization
and channel of the SCOI, assuming an ideal FEC and bit-wise mismatched
decoding optimized for the AWGN channel \cite{alvarado2018achievable,fehenberger2018multiset,agrell2021performance}.

As regards CPR, two different approaches are considered: mean phase
rotation (MPR) and blind phase search (BPS). On the one hand, MPR
is the typical approach employed in simulations---when the laser
phase noise is not considered and, thus, CPR not required---to remove
the (constant in time) expected value of the nonlinear phase rotation
induced by fiber nonlinearity for a given total launch power. In practice,
the MPR is estimated by a simple data-aided procedure---i.e., by
averaging the instantaneous phase rotation experienced by all the
transmitted symbols after propagation---and then removed from
all the received symbols. On the other hand, BPS is a practical CPR
algorithm typically employed with QAM constellations to track the
random fluctuations of the carrier phase induced by laser phase noise
\cite{pfau2009BPS}. In a nutshell, BPS estimates the carrier phase
at discrete time $k$ as the phase rotation (selected among a certain
number of test phases) that minimizes the mean square error between
the rotated symbols and the corresponding QAM decisions over a window
of $2N_{\text{CPR}}+1$ symbols centered at time $k$. A shorter window
can track faster phase variations, whereas a longer window is required
to average out the impact of ASE noise more effectively. The optimal
window width is the trade-off between these two effects. In the following,
we optimize $N_{\text{CPR}}$ numerically and consider 64 test phases
in a $\pi/2$ interval. The MPR is nearly equivalent to a BPS with
$N_{\text{CPR}}\to\infty$. 

The phase estimated by BPS is affected by an ambiguity of multiples
of $\pi/2$ due to the $4$-fold rotational symmetry of conventional
QAM constellations. This ambiguity may induce detrimental cycle slips,
but can be avoided by using, for example, differential coding or pilot
symbols \cite{pfau2009BPS}. In our simulations, for the sake of simplicity,
when laser phase noise is included in the system, we further apply
a supervised cycle-slip compensation after BPS \cite{mello2018interplay}.

\section{Numerical Results\label{sec:System-performance}}

\begin{figure}[tp]
\centering \includegraphics[width=1\columnwidth]{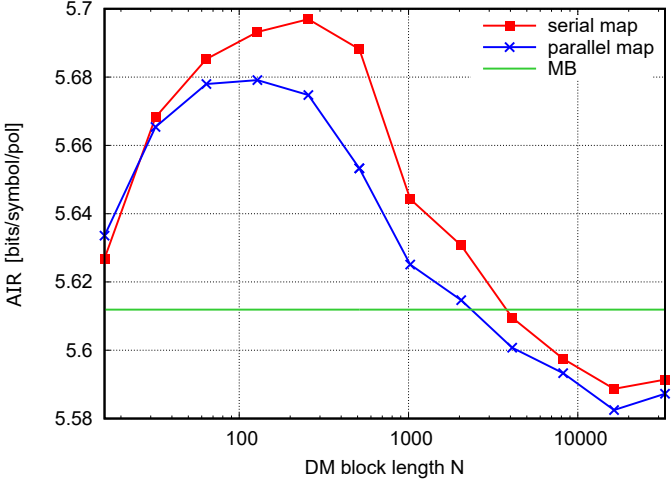}

\caption{\label{fig:GMIvsN_MAP}AIR versus block length for different strategies
to map amplitudes to symbols. $15\times80$km link, EDC, $\Delta\nu=0$~kHz,
no BPS, SS.}
\end{figure}

First, we investigate the performance of the different PAS and bit-to-amplitude
mapping schemes presented in Section~\ref{sec:Probabilistic-amplitude-shaping}.
Figure~\ref{fig:GMIvsN_MAP} compares the performance of the serial
and parallel bit-to-amplitude maps at the optimal launch power for
different block lengths. In this case, SS is used (implemented with
the ESS algorithm), DBP is not applied, the laser linewidth is set
to zero, and the BPS carrier-recovery algorithm is not employed. As
a reference, the performance obtained with MB-distributed i.i.d. symbols
(optimal in the linear regime) is also shown. When the block length
is very short, up to $N\approx32$, the two maps perform the same,
since the performance is limited by the large rate loss. For longer
block lengths, the serial map performs better than the parallel one
and achieves the highest AIR, with a gain of approximately $0.02$~bits/symbol/pol
over the parallel map---a similar behaviour was shown in \cite{fehenberger2019analysis,peng2020transmission}
for a single polarization QAM map. The superiority of the serial map
is explained by the fact that, for a given DM block length, it constraints
the signal energy on a four-time shorter time interval (twice shorter
in the single-polarization case) compared to the parallel map, reducing
the intensity fluctuations and the corresponding nonlinear phase noise.
Both curves improve up to an optimal point ($N\approx256$ and $N\approx128$
for the serial map and for the parallel map, respectively) and decrease
afterwards to approach the MB curve for $N\rightarrow+\infty$. The
peaky behaviour of both curves and the presence of a nonlinear shaping
gain compared to the MB reference curve are analyzed in detail in
the following, where only the serial map is considered, due to its
superior performance.

\begin{figure}[tp]
\centering \hfill{}(a)\hfill{}

\includegraphics[width=1\columnwidth]{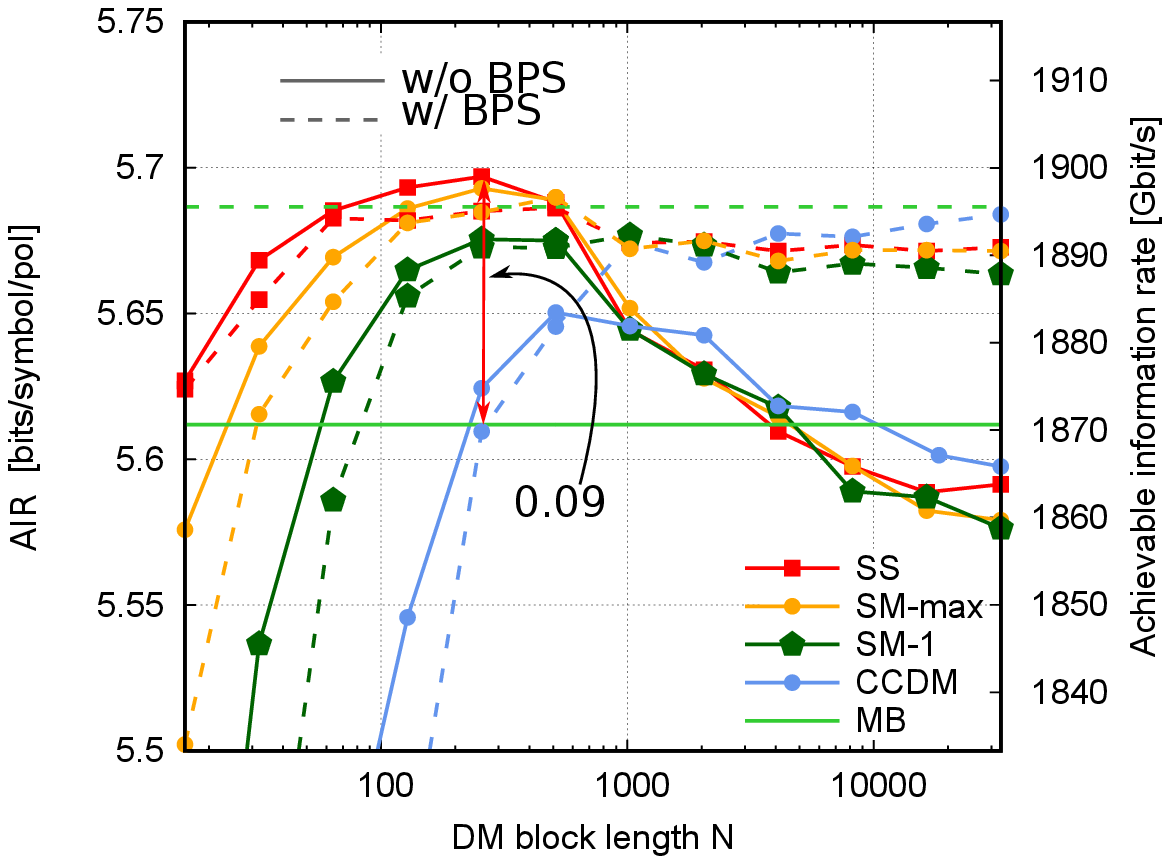}

\hfill{}(b)\hfill{}

\includegraphics[width=1\columnwidth]{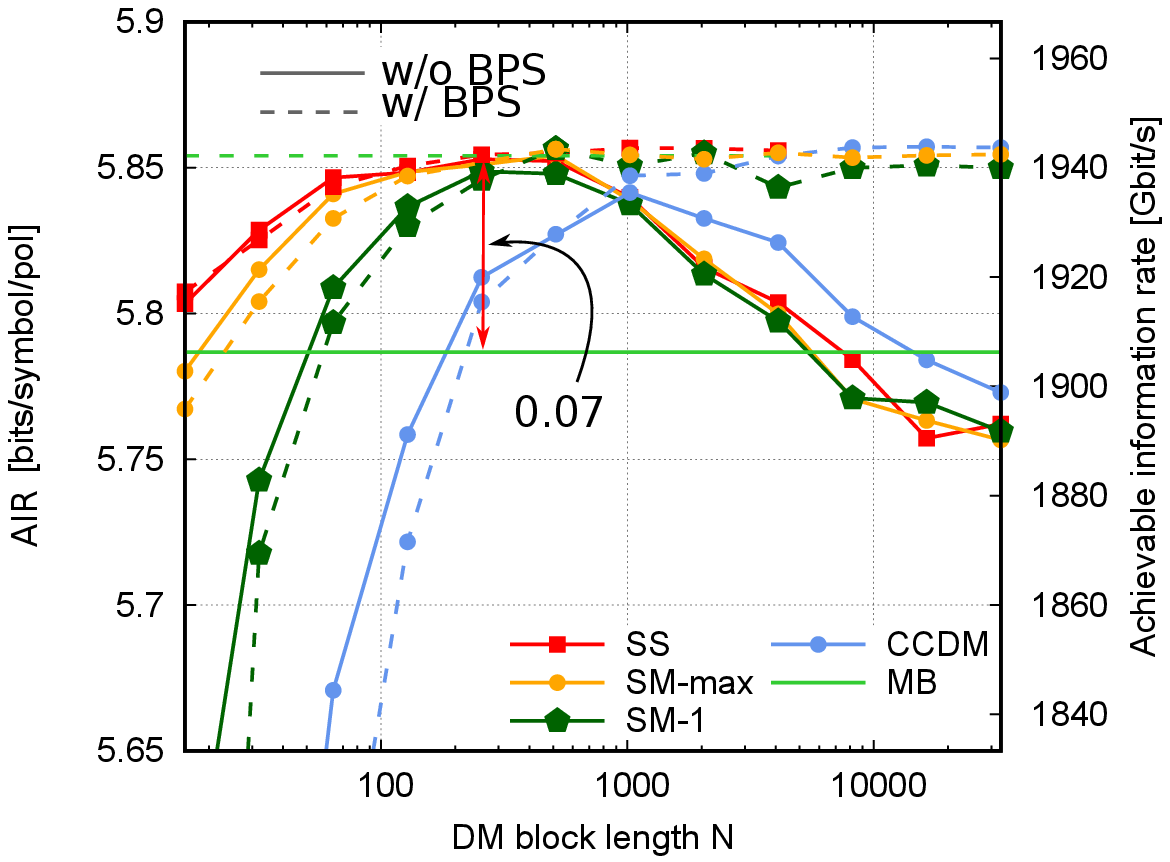}\caption{\label{fig:GMIvsN}Maximum AIR versus DM block length for $15\times80$km
link with different DM techniques and with and without BPS (a) EDC
with $N_{\text{CPR}}=24$, and (b) DBP with $N_{\text{CPR}}=16$.
$\Delta\nu=0$~kHz.}
\end{figure}
The performance of different DMs are compared in Figs.\ \ref{fig:GMIvsN}(a)-(b)
as a function of the DM block length and with (solid lines) or without
(dashed lines) BPS, for the case without DBP in (a) and when ideal
DBP is included in (b).

On the one hand, Fig.\ \ref{fig:GMIvsN}(a) shows that, when BPS
is not employed, the AIR improves up to a certain optimal value of
the block length, after which it decreases again, approaching the
AIR obtained with i.i.d. MB symbols (the distribution obtained when
the DM block length tends to infinity). The difference between the
peak performance and the MB line is the nonlinear shaping gain \cite{fehenberger2016JLT,fehenberger2020mitigating,geller2016shaping}.
This behaviour depends on the combination of two opposing trends:
on the one hand, a longer block length implies a lower rate loss and
hence a better linear performance; on the other hand, it also implies
a weaker correlation between the symbols produced by the DM, whose
intensity fluctuates in time more freely, causing a stronger nonlinear
phase noise. The optimal block length is the trade-off between linear
performance (rate loss) and nonlinear shaping gain (correlations induced
by DM). A similar behavior is observed for all the considered DMs,
and both with or without DBP. However, the nonlinear shaping gain
is the largest for the SS (approximately equal to $0.085$~bit/symbol/pol),
just slightly smaller for SM, smaller for SM-1, and the smallest for
CCDM, following the same ranking shown in the linear regime. Since
SM reduces the intensity fluctuations of the signal with respect to
SS, one could expect the SM or the SM-1 to provide the best nonlinear
performance---as shown in \cite{gultekin2022mitigating} for a $205$~km
fiber. However, the results show that the linear performance prevails.
In fact, a lower rate loss (as for the SS) allows to reduce the DM
block length and, consequently, to enforce a stronger constraint on
the possible intensity fluctuations of the signal, with less nonlinear
phase noise. The nonlinear behavior of the CCDM can be predicted using
the energy dispersion index \cite{wu2021temporal,wu2021EEDI}. The
superior performance of SS with respect to CCDM in the nonlinear regime
was also shown in \cite{amari2019enumerative} and experimentally
in \cite{Nguyen_OFC2022}.

On the other hand, Fig.\ \ref{fig:GMIvsN}(a) shows that when BPS
is employed, the performance of all methods improves almost monotonically
towards the MB curve, which is higher than without BPS. In this case,
the additional nonlinear shaping gain provided by the optimization
of the block length is negligible, meaning that the BPS is mitigating
the same nonlinear phase noise that would be mitigated by short-block-length
PAS. In practice, when the BPS is employed, the optimal performance
can be obtained by using PAS with a sufficiently high block length
to reduce the rate loss, without a specific block length optimization.
In this case, the minimum required block length to achieve the optimal
performance depends on the considered DM, but the optimal performance
does not. The performance for short block lengths follows the same
ranking given in the linear regime, as for the case without BPS. However,
for short block lengths, the performance of the curves with BPS are
slightly worse than those with BPS---this happens because the BPS
(which in this case plays no useful role since the nonlinear phase
noise is completely mitigated by the short-block-length PAS and there
is no laser phase noise) has been optimized for the MB curve, which
has a larger SNR.

Fig.\ \ref{fig:GMIvsN}(b) shows the same as Fig.\ \ref{fig:GMIvsN}(a),
but adding ideal DBP. The performance improves, since DBP is applied,
but the overall behavior does not change with respect to Fig.\ \ref{fig:GMIvsN}(a)---in
fact, even the negligible nonlinear shaping gain observed at intermediate
block lengths in Fig.\ \ref{fig:GMIvsN}(a) vanishes. This result
confirms that short-block-length PAS and BPS mitigate the same nonlinear
effects (mostly the nonlinear phase noise due to inter-channel nonlinearity),
so that the gains they provide, which are similar, do not add up.
On the other hand, DBP and PAS (or DBP and BPS) mitigate different
nonlinearities----intra- and inter-channel, respectively---and their
gains add up.

\begin{figure}[tp]
\centering \hfill{}(a)\hfill{}

\includegraphics[width=1\columnwidth]{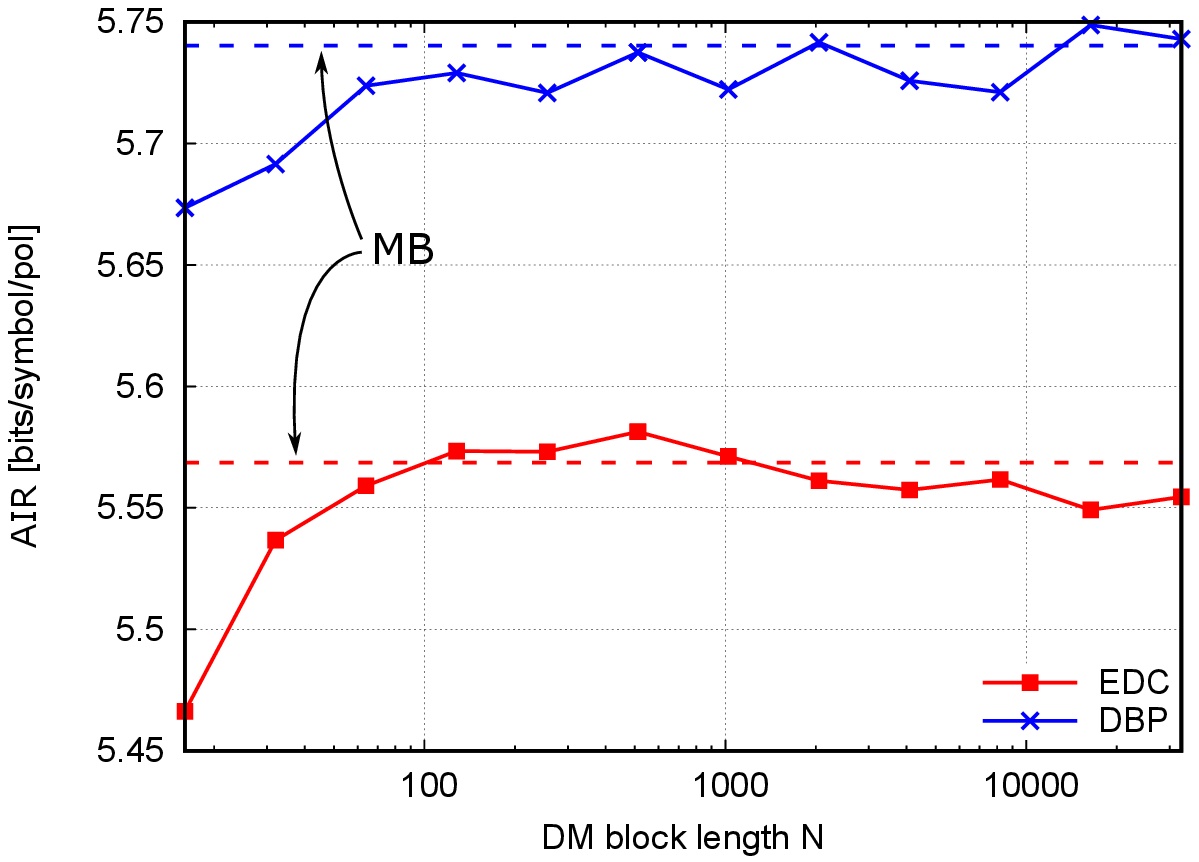}

\hfill{}(b)\hfill{}

\includegraphics[width=1\columnwidth]{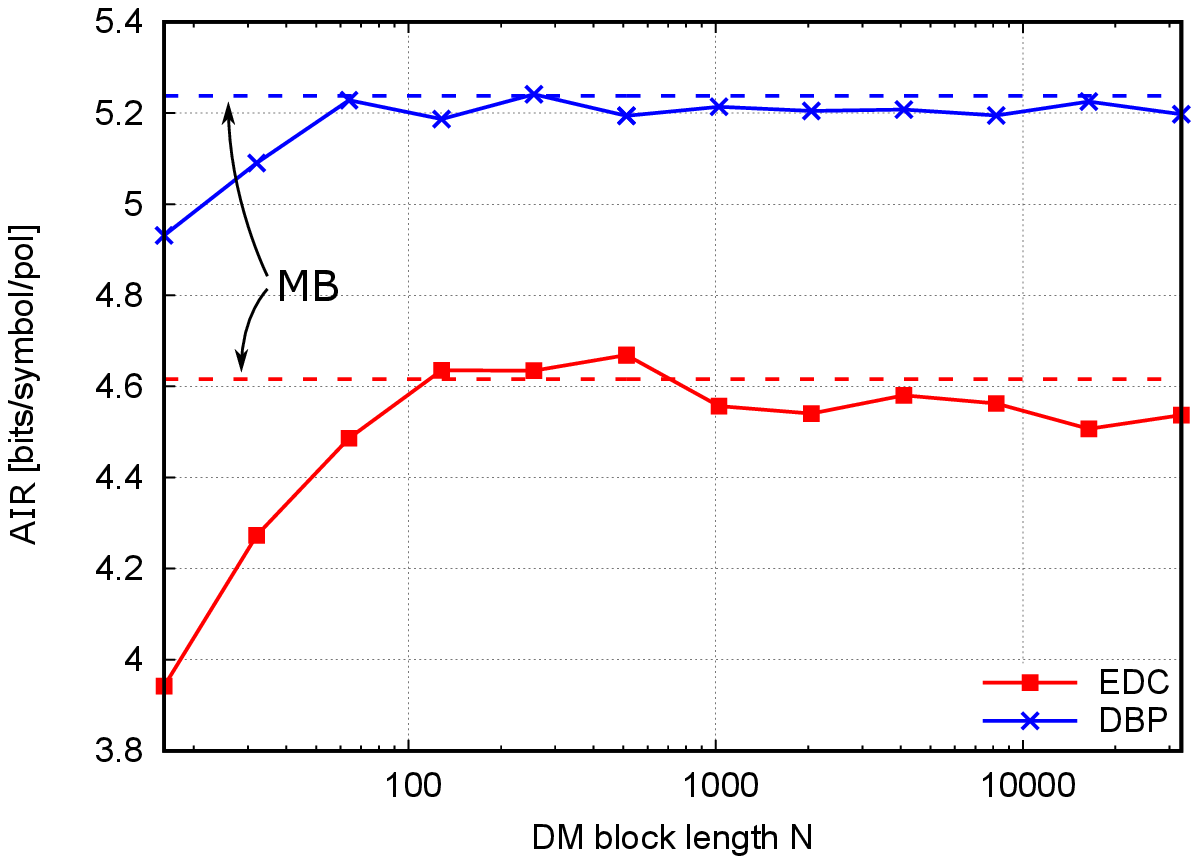}

\caption{\label{fig:GMIvsN_LPN}Maximum AIR versus DM block length using SS
with $\Delta\nu=100$~kHz laser linewidth (a) 15 spans with $N_{\text{CPR}}=38$,
(b) $27$ spans with $N_{\text{CPR}}=92$.}
\end{figure}
An important conclusion that can be drawn from Figs.\ \ref{fig:GMIvsN}(a)-(b)
is that the nonlinear shaping gain provided by PAS is not relevant
when BPS is employed and optimized to minimize nonlinearities. However,
while a CPR algorithm is always included in a system, its time window
(in our case, the $2N_{\text{\ensuremath{\mathrm{CPR}}}}+1$ symbols
over which it estimates the phase) is typically dictated by the laser
linewidth and the system SNR, so that it cannot be freely optimized
to mitigate nonlinear effects. For instance, while nonlinear phase
noise is relatively fast and requires a short time window for its
mitigation, a system with relatively good lasers and low SNR may require
a much longer time window to achieve its optimal performance. The
impact of laser phase noise is investigated in Figs.\ \ref{fig:GMIvsN_LPN}(a)-(b),
which show the AIR versus DM block length for a (a) $15\times80$~km
and (b) $27\times80$~km link, when a laser with linewidth $\Delta\nu=100$~kHz
is considered at the TX and RX sides. The time window of the BPS
algorithm---optimized to mitigate the laser phase noise for the MB
case when DBP is not applied---is $N_{\text{CPR}}=38$ in (a) and
$N_{\text{CPR}}=92$ in (b), the difference being due to the lower
SNR in the second case. In the $15\times80$~km link, the BPS algorithm
has a sufficiently short time window to mitigate most of the nonlinear
phase noise, so that the optimization of the PAS block length does
not yield any additional nonlinear shaping gain with respect to the
case of infinite block length (i.i.d. samples). On the other hand,
in the $27\times80$~km link, the BPS operates on a longer time window
and is not able to mitigate all the nonlinear phase noise---in particular,
the portion that is generated by intrachannel nonlinearity, which
has faster variations. As a result, in this case a small nonlinear
shaping gain of approximately $0.05$~bit/symbol/pol can be observed
when DBP is not employed (note that Fig.\ \ref{fig:GMIvsN_LPN}(a)
and (b) have substantially different vertical scales). A similar behavior---with
a larger gain of $\approx0.1$~bit/symbol/pol---was shown in \cite{civelli2020interplayECOC},
where the laser phase noise was not included.

\begin{figure}[tp]
\centering \hfill{}(a)\hfill{}

\includegraphics[width=1\columnwidth]{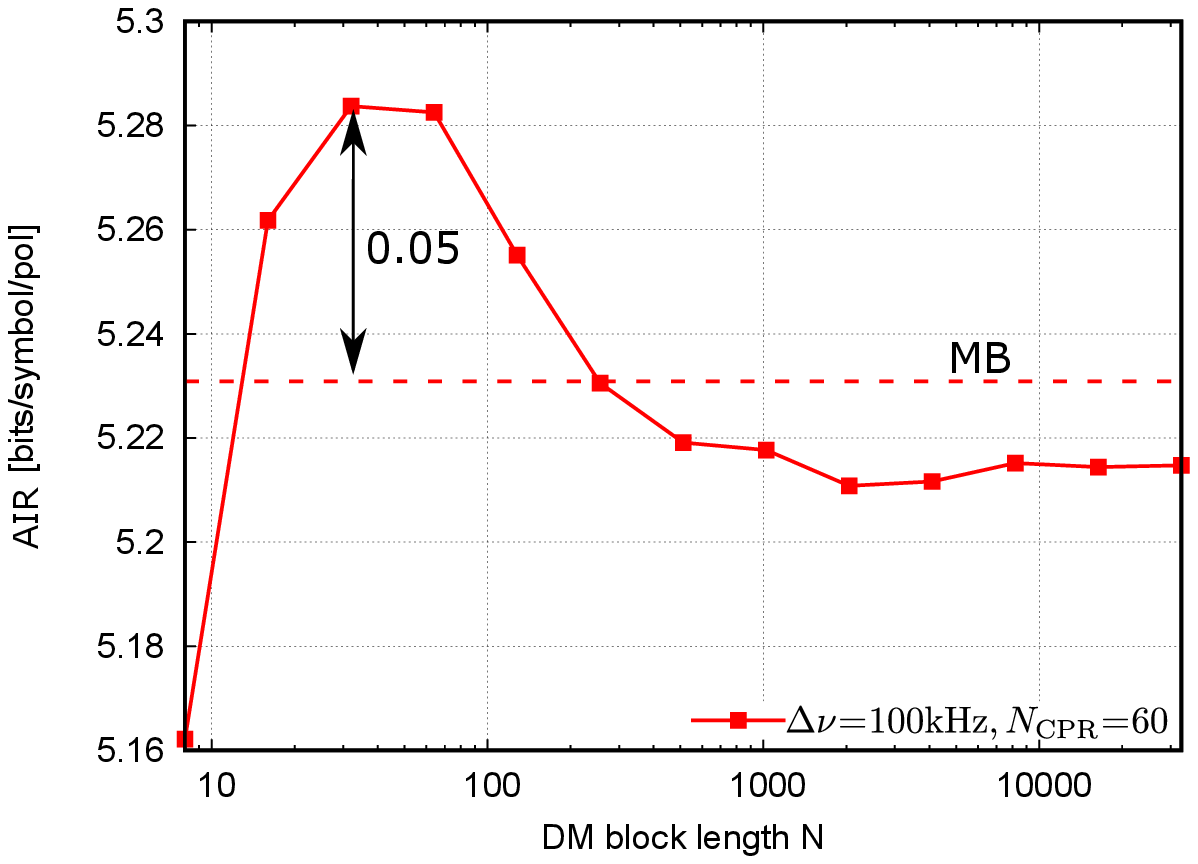}

\hfill{}(b)\hfill{}

\includegraphics[width=1\columnwidth]{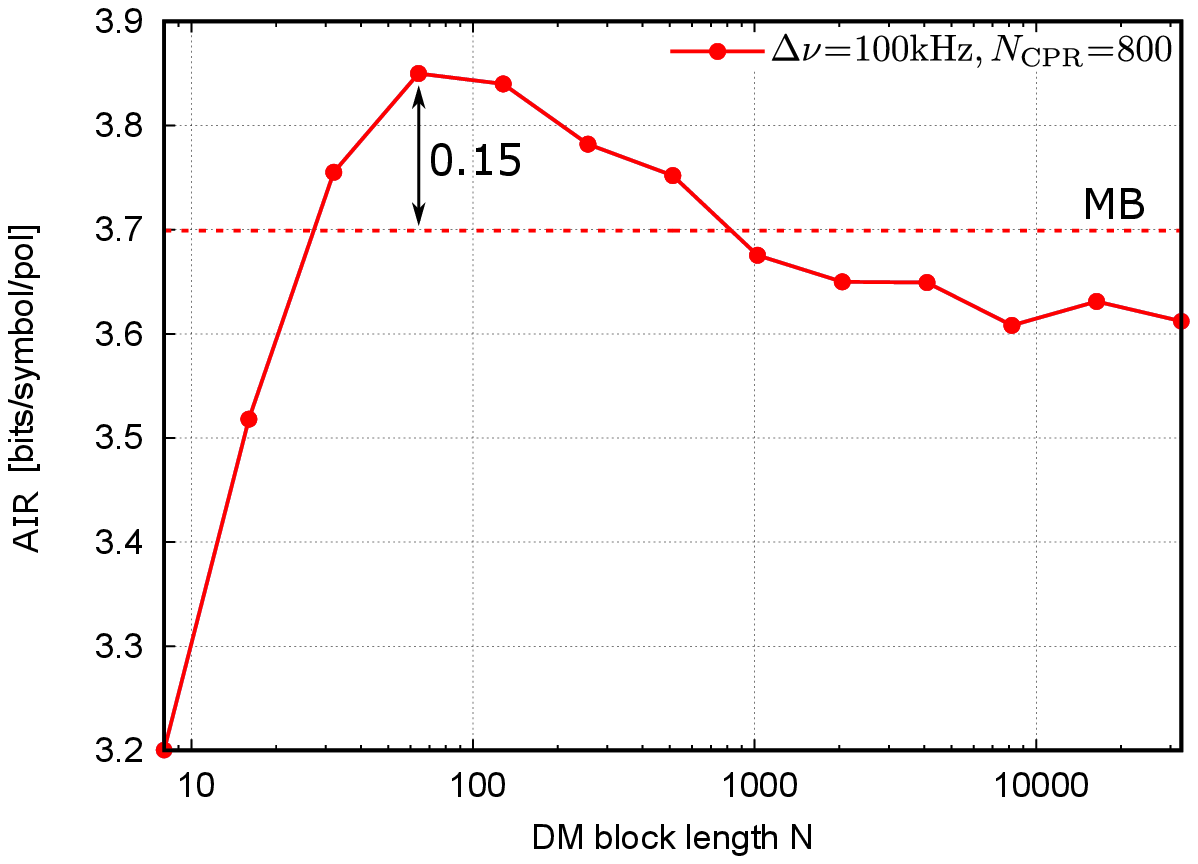}

\caption{\label{fig:GMIvsN_DCF}Maximum AIR versus DM block length using SS
with $\Delta\nu=100$~kHz laser linewidth and EDC, for a (a) $180$\,km
single span of SMF, with $N_{\text{CPR}}=60$, (b) $15\times80$\,km
link with full inline dispersion compensation, with $N_{\text{CPR}}=800$.}
\end{figure}

Next, we consider two rather different links to verify if and how
the behaviour highlighted in the previous cases changes when there
is much less accumulated dispersion. In both cases, we consider SS-based
PAS, EDC at the RX, and we include laser phase noise and the BPS algorithm
with optimized $N_{\mathrm{\text{CPR}}}$. Fig.\ \ref{fig:GMIvsN_DCF}(a)
reports the AIR versus DM block length for a single-span SMF link
of length $180$~km, with $N_{\text{CPR}}=60$. In this case, the
peak AIR is achieved for an optimal block length $N=32$---much shorter
than in previous cases, since the lower accumulated dispersion makes
high-frequency intensity variations more important in the generation
of nonlinear phase noise, reducing the optimal DM block length---with
a gain of approximately $0.05$~bit/symbol/pol with respect to the
ideal case with infinite block length (i.i.d. MB symbols). On the
other hand, Fig.\ \ref{fig:GMIvsN_DCF}(b) reports the AIR versus
DM block length for a $15\times80$~km link with full inline dispersion
compensation, where each 80~km SMF span is followed by $13$\,km
of dispersion compensating fiber (DCF) with $\alpha_{\text{\ensuremath{\mathrm{dB}}}}=0.57$~dB/km,
$\beta_{2}=127.5$\,$\text{ps}^{2}/\text{km}$, and $\gamma=6.5\,\text{W}^{-1}\text{km}^{-1}$.
One additional EDFA with a noise figure of $5$\,dB is added at the
input of each span of DCF, setting the launch power in the DCF $4$~dB
below that in the SMF. In this case, since the SNR is significantly
lower than in the previous cases, the BPS must operate on a longer
time window to average out the impact of noise, and the best performance
is obtained for $N_{\text{CPR}}=800$ symbols. Differently from the
dispersion-unmanaged case, the figure shows that (i) the optimal DM
block length is shorter ($N=64$ rather than $N=256$), due to the
lower accumulated dispersion (as in the single-span case) and (ii)
a large nonlinear shaping gain of $0.15$\,bit/symbol/pol is obtained
even if the BPS is employed, since the long BPS performs similarly
to MPR and is ineffective against nonlinear phase noise.

\begin{figure}[tp]
\centering

\includegraphics[width=1\columnwidth]{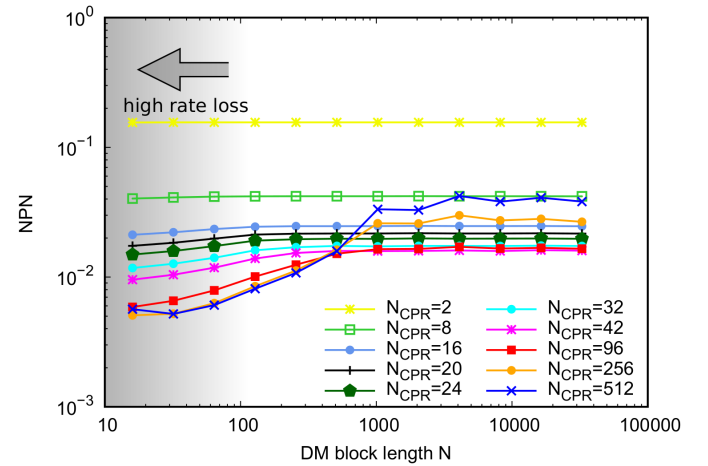}

\caption{\label{fig:FEDI} NPN and EEDI as a function of the DM block length,
for different CPR window lengths.}
\end{figure}

To better understand the interaction between nonlinear shaping gain
and CPR, Fig.\ \ref{fig:FEDI} shows the NPN metric (\ref{eq:RNPN-metric})
as a function of the DM block length and for different values of the
BPS half window $N_{\text{\ensuremath{\mathrm{CPR}}}}$, considering
the same link as in Fig.~\ref{fig:GMIvsN_LPN}(a) (dispersion-unmanaged
$15\times80$\,km without DBP) and the SS strategy for PAS. The metric
is computed for the second channel of the SCOI, $i=2$, and considering
the impact of the $4$ channels of the SCOI. For the sake of simplicity,
we simply set $e_{\text{CPR}}=0.008$ in (\ref{eq:CPRnoise}), which
yields, on average, reasonable results for the considered scenario
and range of $N_{\mathrm{\text{CPR}}}$ values.\footnote{In general, the parameter $e_{\text{CPR}}$ depends on the modulation
format, SNR, and BPS half window $N_{\text{\ensuremath{\mathrm{CPR}}}}$
\cite{pfau2009BPS}. Thus, more accurate results could be obtained
by a detailed characterization of the BPS behavior and dependence
on these parameters. Our simple choice is, however, sufficient to
show the accuracy of the proposed metric.} At the optimal launch power, the (linear) SNR value is $E_{s}/N_{0}=\unit[17]{dB}$.
For a very long BPS window (e.g., $N_{\text{CPR}}=512$), the BPS
is too slow to track the nonlinear phase noise and behaves in practice
as the MPR. In this case, the phase estimate (\ref{eq:theta_estimated})
converges to the average nonlinear phase rotation, the variance of
the CPR noise (\ref{eq:CPRnoise}) vanishes, and the NPN metric (\ref{eq:RNPN-metric})
measures only the amount of generated nonlinear interference. As a
result, the metric behaves similarly to the EEDI: it initially increases
with $N$, until it saturates (for $N\approx1024$) to the value that
would be obtained for i.i.d. MB symbols. In this case, the use of
a relatively short block length ($N<1024)$ is beneficial to reduce
the nonlinear phase noise. However, the reduction of the block length
causes also an increase of the DM rate loss. The combination of these
two effects results in the behavior shown in Fig.~\ref{fig:GMIvsN}(a)
(w/o BPS), with an optimal block length that maximizes the AIR. 
On the other hand, when decreasing the BPS window, the noiseless phase
estimate (\ref{eq:theta_estimated}) becomes more accurate, while
the variance of the CPR noise (\ref{eq:CPRnoise}) increases. In other
words, the BPS becomes faster but more noisy. As a result, the NPN
metric (\ref{eq:RNPN-metric}) decreases for long DM block length,
where the phase noise term dominates, and increases for short DM block
length, where the CPR noise term dominates. Thus, the NPN curves tend
to flatten and the dependence on the DM block length becomes weaker.
This explains why the behavior of the AIR in Fig.~\ref{fig:GMIvsN}(a)
changes when the BPS is included: in this case, using a short DM block
length is no longer beneficial, since the BPS already mitigates the
nonlinear phase noise caused by the intensity fluctuations of the
signal. In fact, to go beyond the mitigation capabilities of the BPS
and see an additional SNR improvement caused by PAS, the DM block
length should be reduced too much (e.g., $N<256$), where the DM rate
loss is however too high. Finally, the use of a too short window (e.g.,
$N_{\mathrm{\text{CPR}}}=2$ or $8$) makes the BPS too noisy, with
a significant performance degradation at any DM block length.

\begin{figure}[tp]
\centering

\hfill{}(a)\hfill{}

\includegraphics[width=1\columnwidth]{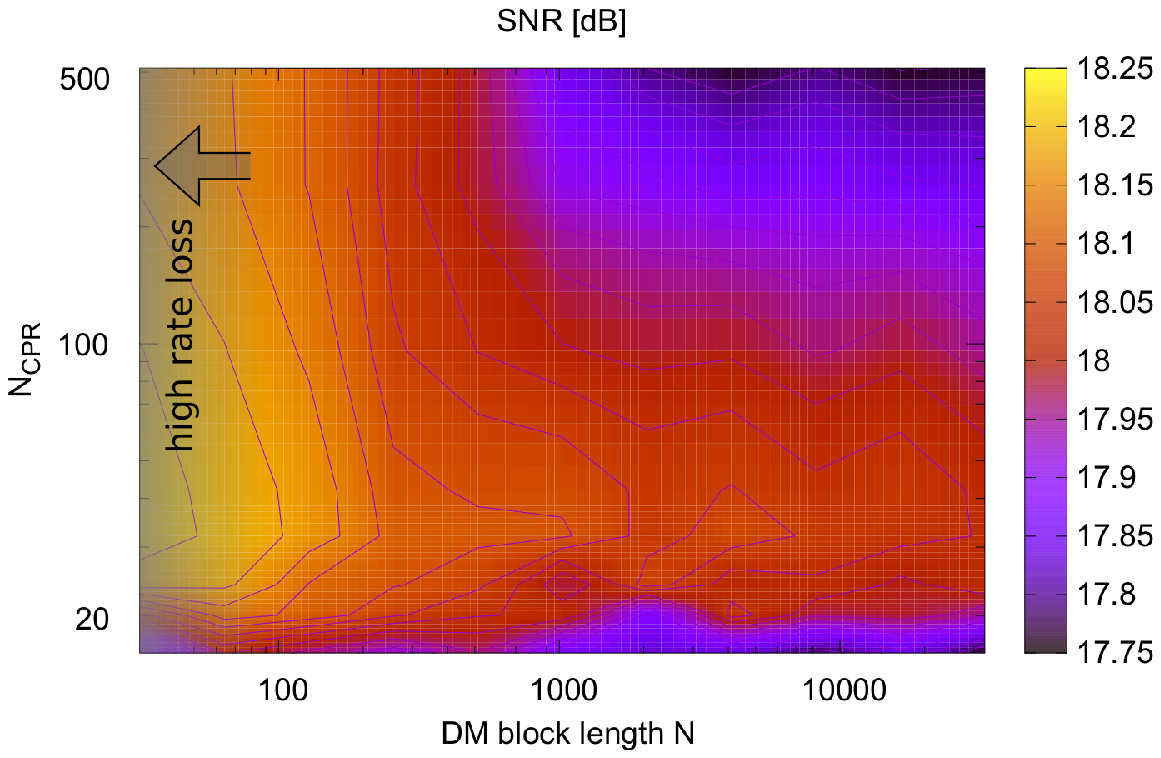}

\hfill{}(b)\hfill{}

\includegraphics[width=1\columnwidth]{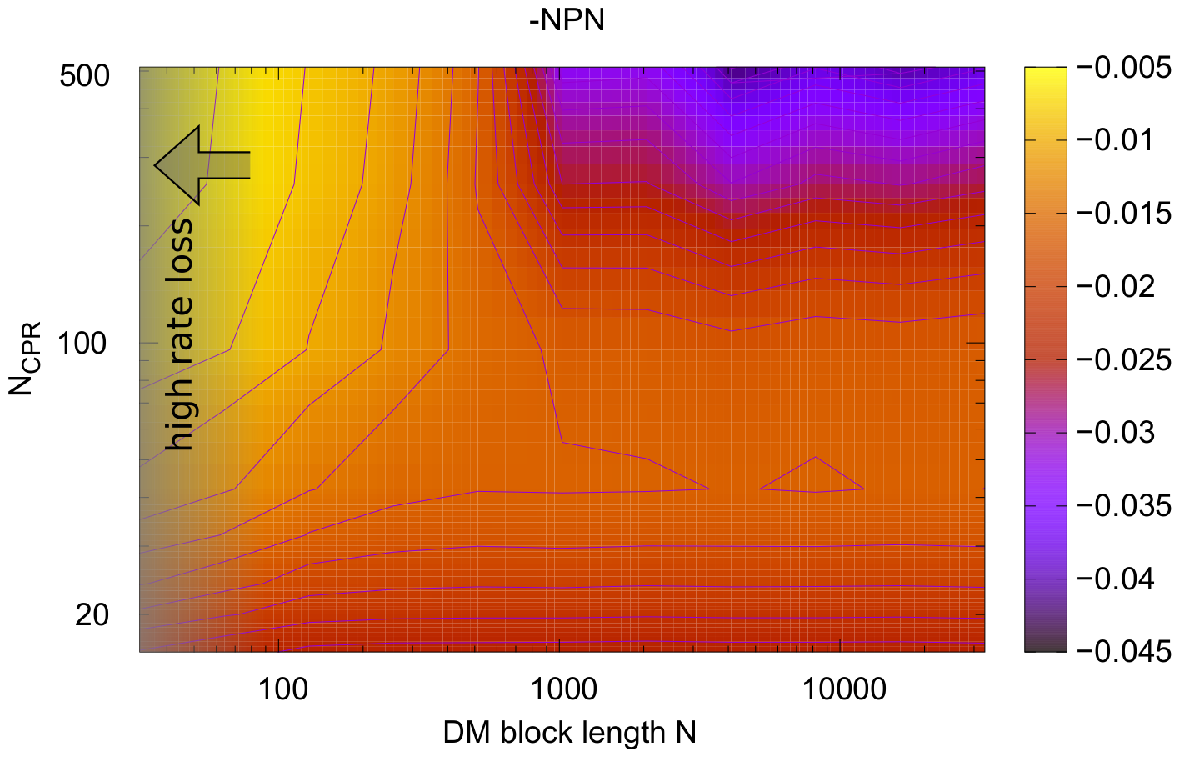}

\hfill{}(c)\hfill{}

\includegraphics[width=1\columnwidth]{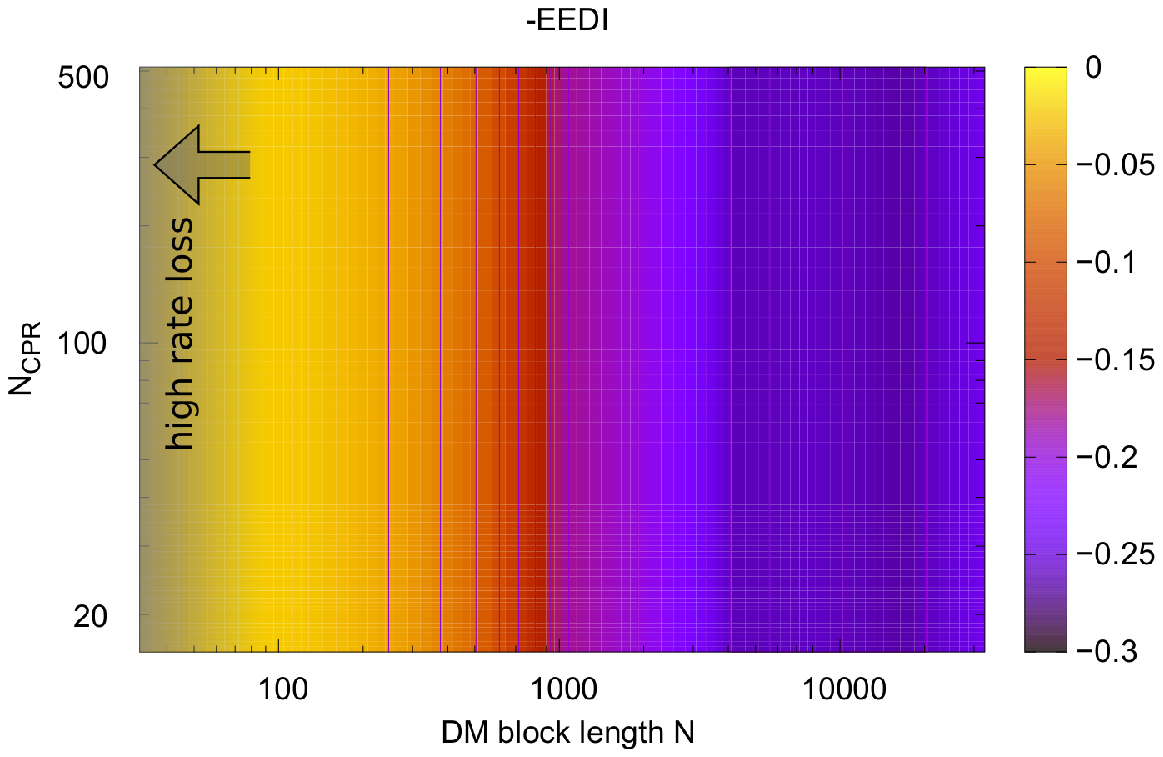}

\caption{\label{fig:FEDI-1}Heat map of (a) simulated SNR (b) $-$NPN and (c)
$-$EEDI as a function of DM block length $N$ and CPR window $N_{\text{CPR}}$.
$15\times80$km link, EDC, $\Delta\nu=\unit[0]{kHz}$.}
\end{figure}

Finally, Figs\ \ref{fig:FEDI-1}(a-c) show the heat maps of SNR,
$-$NPN and $-$EEDI, respectively, as a function of DM block length
$N$ (x-axis) and CPR window $N_{\text{CPR}}$ (y-axis), in the same
setup as Fig.~\ref{fig:GMIvsN_MAP}, i.e., using SS, dispersion compensation
and $\Delta\nu=\unit[0]{kHz}$ laser linewidth, considering the second
channel of the SCOI. The SNR is obtained through extensive numerical
simulations; the NPN metric is evaluated from (\ref{eq:RNPN-metric}),
with $E_{s}/N_{0}=\unit[17]{dB}$, $e_{\text{\ensuremath{\mathrm{CPR}}}}=0.008$,
and accounting for the 4 channels of the SCOI; and the EEDI is evaluated
from (\ref{eq:EEDI}),(\ref{eq:weighted_sum_energies}) with $\lambda=0.985$,
previously optimized (through numerical simulations) to maximize the
correlation with SNR in the case without CPR. Comparing \ref{fig:FEDI-1}(a)
and \ref{fig:FEDI-1}(b), it is evident that SNR and -NPN have a similar
behavior and are highly correlated: (i) when the block length $N$
is very short, they achieve their best values, which are however not
practically useful since associated with a high DM rate loss; (ii)
for larger $N$ values, the best performance is obtained when $N_{\text{CPR}}$
is large enough to average out the noise and small enough to mitigate
nonlinear interference; (iii) with an optimized $N_{\text{CPR}}$,
the performance is almost independent of PAS block length $N$; (iv)
for very large (but suboptimal) $N_{\mathrm{\text{CPR}}}$ values,
BPS behaves as MPR and the dependence of the metric on the DM block
length $N$ becomes more evident, causing the nonlinear shaping gain
observed in these cases. Conversely, the EEDI shown in Fig.\ \ref{fig:FEDI-1}(c)
is independent of $N_{\text{CPR}}$ by definition; therefore, it is
weakly correlated with the SNR and has the correct dependence on the
DM block length only for very large $N_{\mathrm{\text{\ensuremath{\mathrm{CPR}}}}}$
(or, equivalently, when MPR is employed). In fact, the correlation
coefficient between SNR and $-$NPN over the entire range of $N$
and $N_{\mathrm{\text{CPR}}}$ values considered in Fig.~\ref{fig:FEDI-1}
is $0.99$, while the correlation coefficient between SNR and EEDI
is only $0.02$. These results confirm that the proposed NPN metric
is highly correlated with the system SNR and predict accurately its
dependence on the DM block length, even in the presence of CPR.

\section{Discussion and Conclusion\label{sec:Conclusion}}

In this work, we have investigated the performance of different PAS
schemes in the presence of fiber nonlinearity, considering a conventional
WDM setup, different link configurations, and the presence of carrier
phase recovery (CPR). First, we have compared different amplitude-to-symbol
mapping, showing that it is convenient to pack together the amplitudes
produced by a single DM across the four dimensions given by quadratures
and polarizations, hence reducing the intensity fluctuations of the
signal over time.

Next, we have compared different DM implementations---namely, sphere
shaping (SS), shell mapping, and CCDM. In all the considered cases,
increasing the DM block length increases the linear shaping gain (since
the rate loss decreases) but reduces the nonlinear shaping gain (since
the signal intensity can change more freely over time), so that the
optimal performance is obtained at some finite block length. Somewhat
counterintuitively, SS always yields the best performance in terms
of achievable information rate, meaning that its superior linear shaping
gain at short block length more than compensates for its lower effectiveness
in constraining the intensity fluctuations of the signal. In a typical
dispersion-unmanaged WDM scenario, SS with an optimal block length
of 256 amplitudes yields a gain of about 0.1~bit/symbol/polarization
compared to the ideal case of infinite PAS block length (i.i.d. symbols).

After that, we have shown that the presence of a CPR algorithm (e.g.,
BPS) may change the overall picture and the above findings quite significantly,
reducing the nonlinear shaping gain provided by a short-block-length
PAS and making it negligible in most of the scenarios considered in
this work. This is due to the ability of BPS (or similar algorithms)
to mitigate not only the laser phase noise for which it is mainly
employed, but also the nonlinear phase noise caused by fiber nonlinearity.
In this case, reducing the PAS block length brings no additional benefits.
The latter result appears particularly important when considering
that the presence of a CPR algorithm is often neglected in the numerical
investigations that can be found in the literature, but is always
necessary in real systems. The reduction of the nonlinear shaping
gain is more evident in the scenarios where the mitigation of nonlinear
phase noise by BPS is more effective, that is, for higher SNR (which
allows using a shorter BPS window \cite{pfau2009BPS}), more accumulated
dispersion (which increases the coherence time of nonlinear phase
noise \cite{secondini2012analytical}), and/or when DBP is included
(which removes intrachannel nonlinearity, against which BPS is less
effective). By contrast, a significant nonlinear shaping gain can
be still observed in some particular scenarios, such as the link with
full inline dispersion compensation and relatively low SNR considered
in this work, where SS with an optimal block length of 64 amplitudes
yields a gain of about $0.15$\,bit/symbol/pol with respect to the
infinite-block-length PAS, even when an optimized BPS algorithm is
included in the system.

Finally, we have introduced a new NPN metric that explains and predict
quite accurately all the behaviors described above. The metric is
derived from the frequency-resolved logarithmic perturbation \cite{secondini2012analytical,Secondini:JLT2013-AIR}
and gives an analytical approximation of the variance of the residual
(after CPR) nonlinear phase noise generated by the intensity fluctuations
of the signal. In contrast to other existing metrics, such as the
EDI and EEDI, the proposed NPN metric relies on more physical grounds
and contains no adjustable parameters, so that its computation depends
directly on the system configuration and does not require any preliminary
tuning based on extensive simulations. Moreover, the NPN metric accurately
predicts the dependence of system SNR on both the PAS block length
and the size of the CPR window.

In conclusion, the results presented in this work highlight the importance
of including CPR in the analysis and optimization of PAS in the nonlinear
regime. In fact, the dependence of the nonlinear shaping gain on the
specific PAS implementation (DM and amplitude-to-symbol mapping) and
block length---which is observed in many scenarios in the absence
of CPR---may become less relevant in the presence of CPR. In many
cases, different DMs may perform equally well in the nonlinear regime,
provided that a sufficiently long block length is employed, meaning
that other factors (e.g., complexity) may play a more important role
in the design of the system. This is, however, not always true, since
there exist some specific scenarios (e.g, with low SNR and small accumulated
dispersion), where the dependence of nonlinear shaping gain on the
employed DM and block length is still relevant. From a practical point
of view, the NPN metric proposed in this work allows to account for
all these factors accurately, and can be used as a simple guide to
jointly optimize PAS and CPR without performing extensive simulations.

\bibliographystyle{ieeetr}

\end{document}